\documentclass[fleqn,12pt,twoside]{article}
\usepackage{espcrc1}

\usepackage{graphicx}

\usepackage[figuresright]{rotating}

\newcommand{\AmS}{{\protect\the\textfont2
  A\kern-.1667em\lower.5ex\hbox{M}\kern-.125emS}}
\title{Shell effect in Pb isotopes near the proton drip line}
\author{C. Samanta\address[SINP]{Saha Institute of Nuclear Physics,
        1/AF Bidhannagar, Calcutta 700064}\address[VCU]{Physics Department, Virginia 
       Commonwealth 
       University, Richmond, Virginia 23284},
                S. Adhikari\addressmark[SINP]}

\begin{document}
\maketitle

\begin{abstract}
A mass formula (BWM) without shell effect is employed to study the variation 
of the shell effect in Pb isotopes 
through comparison with the experimental data. Unlike other 
macroscopic formulae, the BWM reproduces 
the general trend of the binding energy versus neutron number curves of all the nuclei from 
Li to Bi. The shell effect in Pb-isotopes reduces to $\sim$ 56 keV at
$N=106$ but, increases gradually for $N<106$, indicating increasing 
shell effect in Pb near the proton drip line.\\
\end{abstract}
\vskip 1cm                  
Quenching of the $N=82$ magic
neutron shell below $^{132}$Sn \cite{kr93,ch95} was suggested in response to 
an astrophysical quest to properly reproduce the isotopic solar r-process
abundances in the $A\simeq120$ mass region. As the shell effect is not a directly measurable 
quantity, one tries to estimate it from 
various theoretical prescriptions and looks for its signature through comparison with the 
experimental data. When the normalised mass deviations are plotted, 
the "unquenched" FRDM model \cite{mo95} 
delineates similar deviation from the experimental data for both Cd ($N>78$)
and Pb ($N=105-120$) isotopes, indicating possible quenching
of $N=82$ and $Z=82$ shell gaps. While 
experimental signature of  $N=82$ shell quenching in $^{130}$Cd
is found \cite{ka00,di03}, the $Z=82$ shell quenching still
remains controversial.
Earlier, from the analysis of a few 
$\alpha$-decay experiments it was predicted that
the shell closure effect at $Z=82$ may disappear in the vicinity of 
$N=112-114$ \cite{to84,br92,sc79}. But the observation of large variation 
in the hindrance factors of $l=0$ $\alpha$-decay to the excited 0$^+$ 
state in even-even Po, Pb, Hg and Pt nuclei indicates persistence of the 
$Z=82$ "shell gap" at the neutron-defficient side \cite{wa94}.

The "shell effect"  in nuclei in principle contains the deformation
and  shell closure effects. Extraction of the shell effect through 
comparison of the experimental mass
with the liquid drop mass formula of Bethe-Weizs$\ddot{a}$cker (BW) is
quite well known.  The BW formula, originally designed for medium and
heavy mass nuclei, fails for light nuclei especially, near the drip
lines \cite{sa02}.  The improved liquid drop model (ILDM)
\cite{my66,so03} also fails to give the correct shape of the
binding energy versus neutron number curves of light nuclei.
In search of a more complete macroscopic formula,
we formulated a mass formula, called BWM \cite{sa02},
modifying the asymmetry and the pairing energy terms of BW. 
The  parameters
of BWM were optimised
to fit the gross properties of the binding energy
versus neutron number curves from Li to Bi. Fitting to
such large number of nuclei over a wider mass range puts stringent
constraints on the choice of parameters.  In fact, as both heavy and
light nuclei are fitted with a single set of parameters, BWM delineates
shell effect more accurately than BW and ILDM.
In the modified-Bethe-Weizs$\ddot{a}$cker mass formula (BWM) 
the expression for the binding energy (BE) is \cite{sa02},
\begin{eqnarray}
BE(A,Z)=15.777 A-18.34 A^{2/3}-0.71\frac{Z(Z-1)}{A^{1/3}}-23.21\frac{(A-2Z)^2}
{[A\times(1+e^{-A/17})]}
\nonumber \\
+ (1 - e^{-A/30}) \delta,
\end{eqnarray}
where the term  $\delta$ = +12 A$^{-1/2}$
for even Z-even N nuclei, and -12 A$^{-1/2}$ for odd Z-odd N nuclei and 0
for odd A nuclei. This formula is applicable 
only for the spherical nuclei having negligible shell effects and it shows 
marked deviation for nuclei
with shell effects.
\begin{figure}[htb]
  \begin{center}
  \includegraphics[height=75mm]{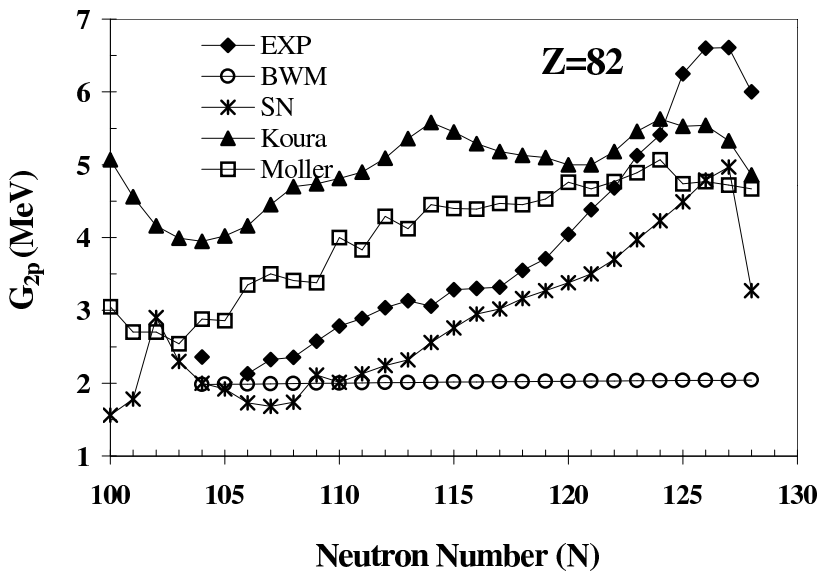}
  \end{center}
  \caption{Variation of G$_{2p}$ with N at Z=82 from experimental data 
\cite{wa00,sc01,no02} and predictions of different mass
formulae (SN \cite{sn83}, Koura \cite{ko00}, M$\ddot{o}$ller \cite{mo95} 
and BWM \cite{sa02}).}
\label{FIG:SAMANTA1}
\end{figure}
\begin{figure}[htb]
\begin{minipage}[t]{75mm}
\begin{center}
  \includegraphics[height=110mm]{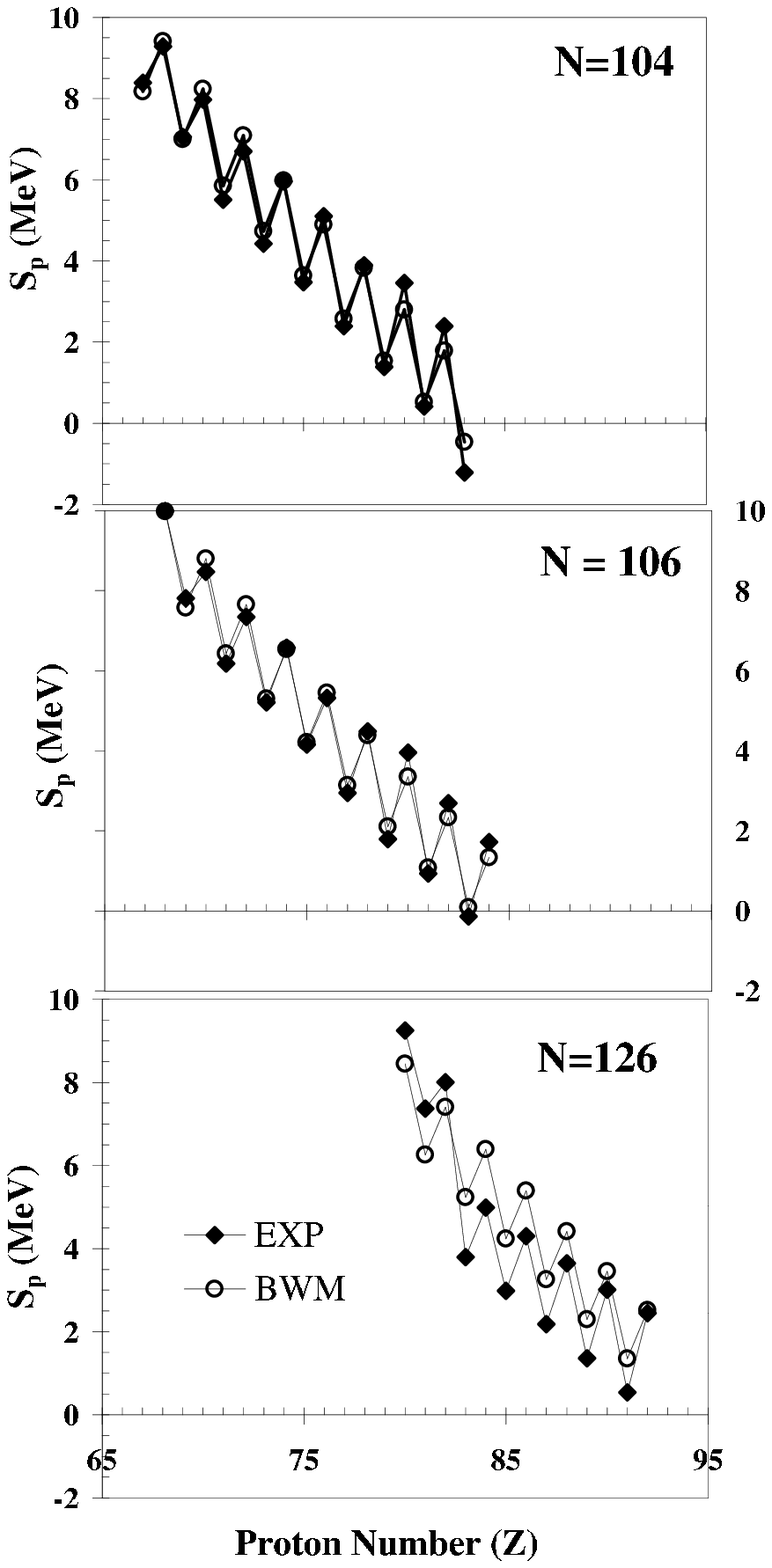}
\end{center}
\caption{Plots of S$_p$ versus Z for N=104, 106 and 126 from experimental
mass data \cite{wa00,sc01,no02} and BWM predictions.}
\label{FIG:SAMANTA2}
\end{minipage}
\hspace{\fill}
\begin{minipage}[t]{75mm}
\begin{center}
  \includegraphics[height=110mm]{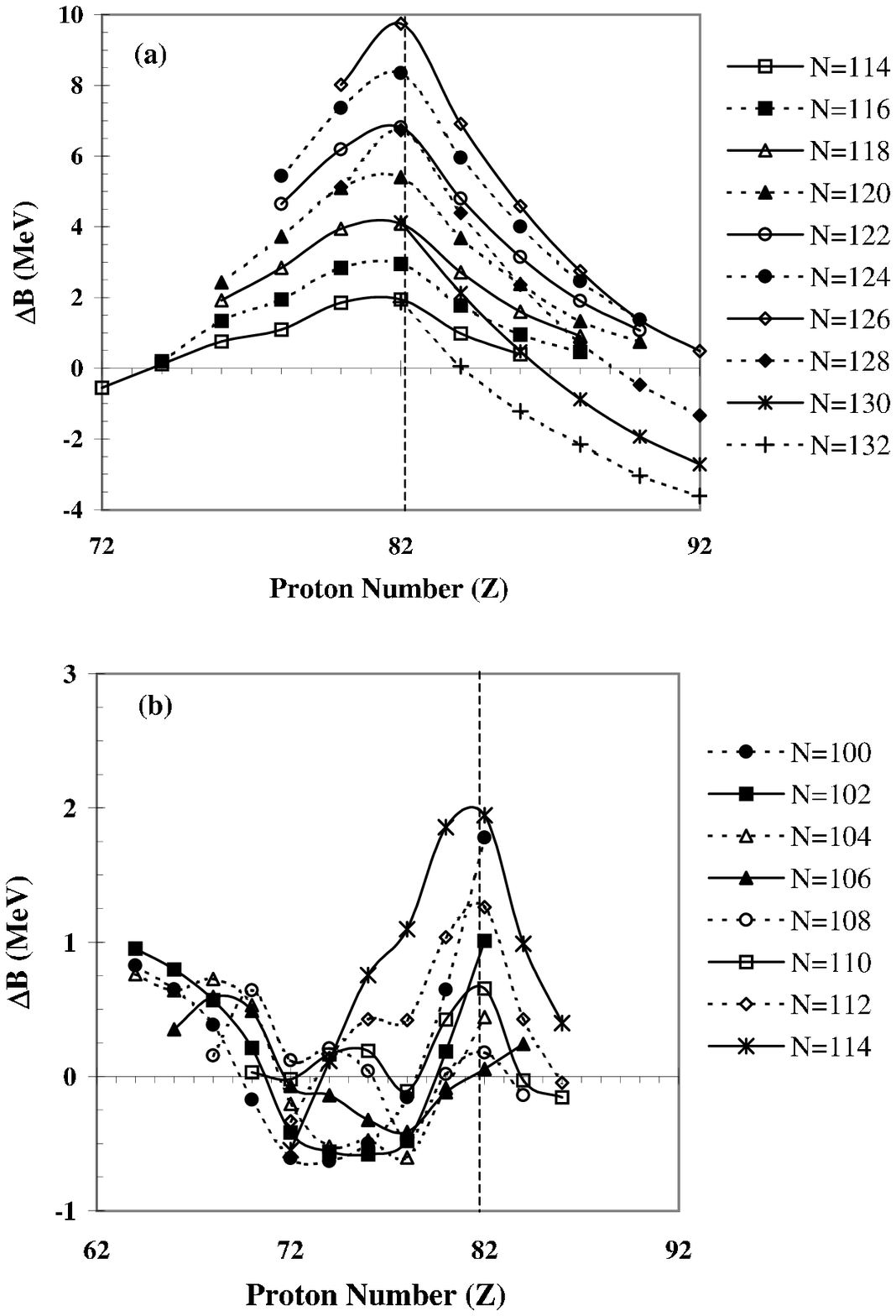}
\end{center}
\caption{Variation of "Shell effect" $\Delta$B with proton number Z for {\bf (a)} N=114-132 and {\bf (b)}
 N=100-114.}
\label{FIG:SAMANTA3}
\end{minipage}
\end{figure}

Figure~\ref{FIG:SAMANTA1} shows plot of
G$_{2p}$ = 2$\times$BE(A,Z) - BE(A-2,Z-2) - BE(A+2,Z+2) for Z=82 computed from
the masses of Hg, Pb and Po isotopes \cite{wa00,sc01,no02}. The
BWM having no shell effect gives a smooth straight line. At $N=106$ 
the difference between the
BWM and experimental data almost disappears. The difference rises again at $N<106$. 
Bender et al. \cite{be02} argued that G$_{2p}$ does not reflect the
$Z=82$ shell gap as, 
most of the Hg and Po
nuclei have deformed ground states. Interestingly, none of the 
"unquenched" mass formulae of refs. 
\cite{mo95,ko00,sn83} can explain the actual shape of the experimental data 
(Figure~\ref{FIG:SAMANTA1})
although they implicitly contain deformation and shell effects. 

Figure~\ref{FIG:SAMANTA2} 
shows the single-proton separation energy (S$_p$) 
from Pb and nearby elements.  
For $Z=82$, a large discrepency between 
the experimental value and the BWM prediction is seen at  $N=126$, as expected. The discrepancy 
reduces greatly at $N=106$ but increases again at $N=104$ where the break at Z=82 reappears 
in the experimental data.
 
In Figure~\ref{FIG:SAMANTA3}, 
the "shell effect" $\Delta$B  i.e., the difference between the binding
energies computed from the experimental masses 
and the BWM is plotted for several nuclei around $Z=82$.
The shell effect at $Z=82$ reduces for both $N<126$ and $N>126$.
Interestingly, after reaching to a low value 
($\sim$56 keV) at $N=106$, the shell
effect increases again for $N<106$. This rise occurs due to mutual support of 
magicity coming from the approaching $N=82$ magic number.

In summary, the G$_{2p}$, S$_p$ and  $\Delta$B
values are computed from the experimental mass defect data and 
a mass formula (BWM) which has no shell effect incorporated \cite{sa02}. 
As the shell effect in a nucleus decreases, 
the discrepancy between the experimental data and the 
predictions of BWM diminishes.
For all the above three quantities a  close agreement between the experimental 
data and the BWM predictions is observed for Pb isotope with N=106. 
Unlike the  G$_{2p}$ and S$_p$ values 
the "shell effect" $\Delta$B computed here 
is not affected by the possible deformations of nearby elements. 

As all the even-even isotopes of 
Pb have spherical ground states, one of the reasons for reduction of shell 
effect near $N=106$ and its increase approaching the proton drip line 
might be the 
change in the $Z=82$ shell closure effect. 
From a self consistent calculation
Bender et al \cite{be02} also found an increase 
of $Z=82$ shell gap approaching the proton drip line, but
the $Z=82$ shell gap was shown to remain large all along.
 It is pertinent to note that
in moving away from the neutron shell closure at $N=126$, the number of 
valence neutrons increases up to the mid-shell configuration and this brings in 
an important energy correlation that contributes to nuclear masses. The large 
amount of valence neutrons will induce a polarization of the closed proton core.
This is clearly present in the mass behaviour of intruder states and should also have important effects on the binding energy \cite{fo03}. 
In a microscopic theory these specific and important correlations should first be removed 
in order to be able to deduce results on a 
changing spherical $Z=82$ shell gap, but such a microscopic calculation
is beyond the scope of this work. 

On the experimental side,
measurements of knockout and pickup cross sections from light Pb isotopes are needed to 
extract the spectroscopic factors and to confirm
either the quenching or, the persistence of the $Z=82$ shell gap.\\


\begin{thebibliography}{99}
\bibitem{kr93} K.L.Kratz et al., Ap. J. {\bf 403},(1993) 216
\bibitem{ch95} B.Chen et al, Phys. Lett. {\bf B355}, (1995) 37
\bibitem{mo95} P.M$\ddot{o}$ller et al., At. Data Nucl. Data Tables {\bf 59}, 
(1995) 185
\bibitem{ka00} T.Kautzsch et al., Eur. Phys. J. {\bf A9}, (2001) 201
\bibitem{di03} I. Dillmann et al., Phys. Rev. Lett., {\bf 91}, (2003) 162503
\bibitem{to84} K.Toth et al., Phys. Rev. Lett. {\bf 53}, (1984) 1623
\bibitem{br92} B.A.Brown, Phys. Rev. {\bf C46}, (1992) 811
\bibitem{sc79} K.H.Schmidt, W.Faust and H.M$\ddot{u}$nzenberg, Nucl. Phys.
{\bf A318}, (1979) 253
\bibitem{wa94} J.Wauters et al., Phys. Rev. Lett. {\bf 72}, (1994) 1329
\bibitem{sa02} C.Samanta and S.Adhikari, Phys.Rev.{\bf C65}, (2002) 037301
\bibitem{my66} W.D.Myers and W.J.Swiatecki, Nucl. Phys. {\bf 81}, (1966) 1
\bibitem{so03} S.R.Souza et al., Phys.Rev.{\bf C67}, (2003) 051602(R)
\bibitem{wa00} Nuclear Wallet Cards, Brookhaven National Laboratory, 2000
\bibitem{sc01} S.Schwarz  et al., Nucl. Phys. {\bf A693}, (2001) 533
\bibitem{no02} Yu.N.Novikov et al., Nucl. Phys. {\bf A697}, (2002) 92
\bibitem{be02} M.Bender et al.,Eur. Phys. J. {\bf A14}, (2002) 23
\bibitem{ko00} H.Koura, M.Uno, T.Tachibana and M.Yamada, Nucl. Phys.
{\bf A674}, (2000) 47
\bibitem{sn83} L.Satpathy and R.Nayak, Phys. Rev. Lett. {\bf 51}, (1983) 1243
\bibitem{fo03} R.Fossion et al., Phys.Rev. {\bf C67}, (2003) 024306; see references therein
\end{thebibliography}
\end{document}